\documentclass[10pt,a4paper]{article}

\pdfoutput=1

\usepackage{jheppub}
\usepackage{amsfonts}
\usepackage{amsmath}
\usepackage{amssymb}
\usepackage{bm}
\usepackage{dcolumn}
\usepackage{epsfig}
\usepackage{graphicx}
\usepackage{graphics}
\usepackage[latin1]{inputenc}
\usepackage{latexsym}
\usepackage{rotating}
\usepackage{hyperref}
\usepackage{dsfont}

\usepackage{color}
\usepackage[active]{srcltx}
\usepackage{amsmath,amsfonts,amssymb,amsthm,amstext,amscd,eucal,srcltx}
\usepackage{epsfig,graphicx,bm}
\usepackage{epstopdf, epsf}
\usepackage{dcolumn}
\usepackage{hyperref}

\newcommand{\bea}{\begin{eqnarray}}
\newcommand{\eea}{\end{eqnarray}}
\renewcommand{\(}{\left(}
\renewcommand{\)}{\right)}

\def\d{{\rm d}}

\begin{document}

\vspace*{3mm}
\title{Klein-Gordon field from the XXZ Heisenberg model}

\author[a]{Jakub Bilski,}

\author[a]{Suddhasattwa Brahma,}

\author[a]{Antonino Marcian\`o,}

\author[b]{and Jakub Mielczarek}

\affiliation[a]{Department of Physics \& Center for Field Theory and Particle Physics, 
Fudan University, 200433 Shanghai, China}

\affiliation[b]{Institute of Physics, Jagiellonian University, \L ojasiewicza 11, 
30-348 Cracow, Poland, EU}

\date{\today}

\abstract{We examine the recently introduced idea of Spin-Field Correspondence focusing on the example of the spin system described by the XXZ Heisenberg model with external magnetic field. The Hamiltonian of the resulting nonlinear scalar field theory is derived for arbitrary value of the anisotropy parameter $\Delta$. We show that the linear scalar field theory is reconstructed in the large spin limit. For $\Delta=1$ a non-relativistic scalar field theory satisfying the Born reciprocity principle is recovered. As expected, for the vanishing anisotropy parameter $\Delta \rightarrow 0$ the standard relativistic Klein-Gordon field is obtained. Various aspects of the obtained class of the scalar fields are studied, including the fate of the relativistic symmetries and the properties of the emerging interaction terms. We show that, in a certain limit, the so-called polymer quantisation of the field variables is recovered. This and other discussed properties suggest a possible relevance of the considered framework in the context of quantum gravity.}

\maketitle

\section{Introduction}
\noindent \label{sec:intro}
Similarities between condensed matters systems and field theories are clearly visible. In both cases, the excitations of the low temperature state are described by particle-like modes characterised by a certain dispersion relation. Both condensed matter systems and field theories are systems characterised by a huge number of degrees of freedom\footnote{In the case of continuous field theories, this number is divergent.}. Furthermore, numerous effects which were considered first in the domain of fundamental field theories found later their counterpart in the analog condensed matter systems (e.g. Hawking radiation). Such a relation has been especially broadly considered in the content of analog condensed matter models of classical and quantum gravity \cite{Barcelo:2005fc}.   

We may then ask whether there is a deeper side of the analogy between field theories and condensed matters systems. In particular, we want to understand whether field theories considered in theoretical physics can be recognized as an approximate description to some discrete condensed matter system such as spin systems. 

Searching for such an identification between fundamental fields and corresponding spin systems is not a new concept. Numerous approaches have been considered in this context in the literature. Notably, from the condensed matter theory side investigations on the classification of topological order thorough \emph{string-nets} \cite{LW} allowed to reconstruct gauge field structures \cite{WenBook}, and to propose to unravelling too the very same concept of elementary particles by reproducing their quantum numbers and dynamical properties \cite{Gu:2010na,Gu:2014wwa}. From the quantum gravity side, the celebrated AdS/CFT correspondence \cite{Maldacena:1997re,Hawking:2000da} is the first example that pops up to our mind, for the sizable amount of studies devoted to this topic in the literature.  

In this article we develop a novel approach to the issue of expressing spin systems in terms of field theories, based on the recently introduced concept of \emph{Spin-Field Correspondence} (SFC) \cite{Mielczarek:2016xql}. The approach emerged as a result of consideration of nonlinear, in particular compact, field phase spaces \cite{Mielczarek:2016rax}. 

\begin{figure}[ht!]
\centering
\includegraphics[width=12cm, angle = 0]{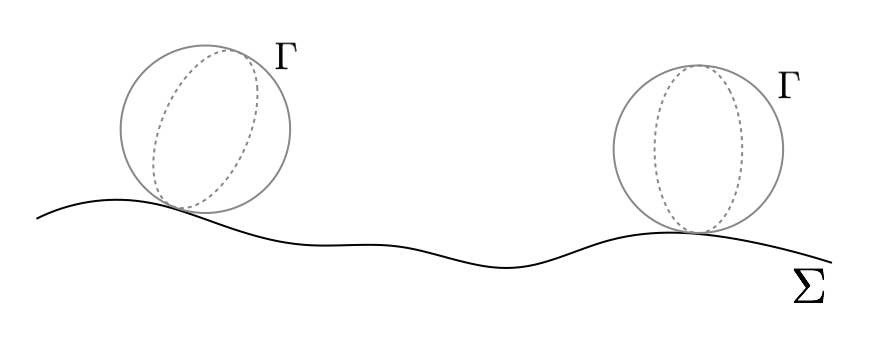}
\caption{Pictorial representation of scalar field spherical phase spaces $\Gamma$ at two different points of the spatial manifold $\Sigma$.}
\label{Fig1}
\end{figure}

Usually, in the field theories, the case of flat phase spaces is considered. Such phase spaces emerge as a semiclassical description of the quantum systems characterized by infinitely dimensional Hilbert space. However, one can speculate that the infinite dimensionality is only an approximation of finite dimensional quantum systems. The quantum systems with finite dimensional Hilbert spaces lead, at the semiclassical level, to the \emph{compact phase spaces}\footnote{If there is no boundary, non-compact phase spaces with finite area are also possible.}. This is due to the fact  that each linearly independent state in the Hilbert space occupies $2\pi \hbar$ area of the phase space. Consequently, the $n$-dimensional Hilbert space leads to a phase space with the area equal to $2\pi \hbar n$. An example of a compact phase space on which we are going to focus here is the spherical phase space. In such a case the phase space of a scalar field at every point of a spatial manifold $\Sigma$ is $\Gamma=S^2$, as schematically depicted in Fig. \ref{Fig1}.

The organisation of the article is the following. In Sec.~\ref{Sec_SFC} the idea of Spin-Field Correspondence for the case of a scalar field is reviewed and definitions necessary for the further part of the article are introduced. Then, in Sec.~\ref{Sec_XYZ} the special case of the XYZ Heisenberg model is considered. It is shown, that by virtue of the Spin-Field correspondence the model leads to the relativistic massive scalar field theory if an appropriate choice of coupling constants is performed. For the the spacial case, which is the  XXZ model characterised by the anisotropy parameter $\Delta$, the relativistic limit is recovered for $\Delta \rightarrow 0$, while for  $\Delta \neq 0$ departures from the relativistic symmetry are expected. The nature of the resulting deformations of the Lorentz symmetry are investigated in Sec.~\ref{Sec_Deformations}. Subsequently, in Sec.~\ref{Sec_Polymer} a step towards a possible relevance of the discussed framework in the context of quantum gravity is performed. In particular, we show that the considered compact field phase space can be reduced to the case of polymerisation discussed in the context of the quantisation of the gravitational degrees of freedom. The results are summarised is Sec.~\ref{Sec_Summary}, where we also outlook some next steps in the development of the presented research direction. 

\section{Spin-Field Correspondence} \label{Sec_SFC}

The idea of Spin-Field Correspondence (SFC) has been introduced employing the fact 
that the phase space of classical angular momentum (spin) is a sphere. This relationship 
is easy to notice by considering rotational invariance of the spin which is satisfied by the 
structure of a sphere. More formally, the issue is expressed by the so-called Kirillov orbit 
methods \cite{Kirillov}, which state that: if the phase space of a classical mechanical 
system being invariant under the action of a group $G$ is an orbit, than at the quantum level, the system is described by irreducible representation of $G$. In case of the spin the group is $G=SU(2)$ for which the orbit $SU(2)/U(1) = S^2$.

Since the phase space is a symplectic manifold, it has to be equipped with the 
closed symplectic form
\begin{equation}
\omega=S\sin\theta\,d\phi\wedge d\theta,
\label{symplectic}
\end{equation}
where $S$ is an absolute value of spin introduced due to dimensional reasons. The total
area of the phase space is $\text{Ar}_{S^2}= \int_{S^2} \omega = 4\pi S$. The $\phi$ and
$\theta$ are the standard angular variables of the spherical coordinate system. For
our purpose, it is convenient to introduce the following change of variables:
\begin{align}
\phi&=\frac{\varphi}{R_1}
,\qquad
\phi\in(-\pi,\pi] \label{coord1}
\\
\theta&=\frac{\pi}{2}-\frac{\pi_{\varphi}}{R_2}
,\qquad
\theta\in(0,\pi)
\label{coord2}
\end{align}
where $\varphi$ and $\pi_{\varphi}$ are our new coordinates. The $R_1$ and $R_2$ and
constants introduced due to dimensional reason. By applying the coordinate change
(\ref{coord1}) and (\ref{coord2}) to the symplectic form (\ref{symplectic}) we find
\begin{equation}
\omega= \frac{S}{R_1R_2} \cos\Big(\frac{\pi_{\varphi}}{R_2}\Big)d\pi_{\varphi}\wedge d\varphi.
\label{symplectic2}
\end{equation}
The symplectic form (\ref{symplectic2}) reduces to the standard Darboux form
($\omega=d\pi_{\varphi}\wedge d\varphi$) in the limit of large $S$, if the following condition 
is satisfied:
\begin{equation}
R_1R_2=S.
\label{normalization}
\end{equation}

By inverting the symplectic form (\ref{symplectic2}), together with the normalisation (\ref{normalization}),
the Poisson bracket can be defined:
\begin{equation}
\{f,g\} :=(\omega^{-1})^{ij}(\partial_i f)(\partial_j g) = \frac{1}{\cos(\pi_{\varphi}/R_2)}
\left( \frac{\partial f}{\partial \varphi} \frac{\partial g}{\partial \pi_{\varphi}}-\frac{\partial f}{\partial \pi_{\varphi}} \frac{\partial g}{\partial \varphi}\right).
\label{Poisson}
\end{equation}
From here, the Poisson bracket between the canonical variables  $\varphi$ and $\pi_{\varphi}$ is found to be
\begin{equation}
\{\varphi,\pi_{\varphi}\} = \frac{1}{\cos(\pi_{\varphi}/R_2)},
\label{Poisson}
\end{equation}
which reduces to the standard case in the $R_2 \rightarrow \infty$ limit.

One can now see explicitly that a sphere is the phase space of a spin. For this purpose let us consider the components of the angular momentum vector $\vec{S} =(S_x,S_y,S_z)$:
\begin{align}
S_x:=&\ S \sin\theta \cos\phi = S \cos \left( \frac{\pi_{\varphi}}{R_2} \right) \cos \left( \frac{\varphi}{R_1} \right),
\label{Sx} \\
S_y:=&\ S \sin\theta \sin\phi = S \cos \left( \frac{\pi_{\varphi}}{R_2} \right) \sin \left( \frac{\varphi}{R_1} \right),
\label{Sy}\\
S_z:=&\ S \cos\theta = S \sin \left( \frac{\pi_{\varphi}}{R_2} \right),
\label{Sz}
\end{align}
fulfilling the condition $S_x^2+S_y^2+S_x^2=S^2$. Using the Poisson bracket (\ref{Poisson}) one can show that the components (\ref{Sx})-(\ref{Sz}) satisfy the $su(2)$ algebra:
\begin{equation}
\{S_x,S_y\} = S_z,  \ \ \ \{S_z,S_x\} = S_y,  \ \ \ \{S_y,S_z\} = S_x.
\end{equation}

The discussion we presented so far concerned a single spin case. However, generalisation to discrete or continuous spin distribution is straightforward \cite{Mielczarek:2016xql}. In the case of continuous spin distribution, the spin vector $\vec{S}$ becomes a function of position vector $\vec{x}$. Consequently, the same concerns the variables $\varphi$ and $\pi_{\varphi}$. Furthermore, one has to assume whether or not the total value of spin $S$ is a nontrivial function of $\vec{x}$. The simplest possibility is that $S$ is spatially constant, which is a natural choice for most of the physical situations\footnote{The case in which $S$ is a function of $\vec{x}$ leads to the spatial dependence of the coupling constants of the resulting scalar field.}. In this case, the Poisson bracket (\ref{Poisson}) generalises to:
\begin{equation}
\{f(\vec{x}), g(\vec{y}) \} = \int \frac{d^3 z }{\cos(\pi_{\varphi}(\vec{z} ) /R_2)}
\left( \frac{\delta f(\vec{x})}{\delta \varphi(\vec{z})} \frac{\delta g(\vec{y})}{\delta
\pi_{\varphi}(\vec{z}) }-\frac{\delta f(\vec{x})}{\delta \pi_{\varphi}(\vec{z})} \frac{\delta g(\vec{y})}{\delta \varphi(\vec{z} ) }\right),
\label{PoissonField}
\end{equation}
which defines the kinematics. In particular, the Poisson bracket between the canonically
conjugated field variables $\varphi$ and $\pi_{\varphi}$ is now  
\begin{equation}
\{\varphi (\vec{x}),\pi_{\varphi} (\vec{y}) \} = 
\frac{\delta^{(3)}(\vec{x}-\vec{y})}{\cos(\pi_{\varphi}(\vec{x}) /R_2)}. 
\label{PoissonField}
\end{equation}
In the $R_2 \rightarrow \infty$ limit the Poisson bracket for the standard scalar field 
theory is recovered. 

\section{Scalar Field Theory from XYZ model} \label{Sec_XYZ}

In Ref.~\cite{Mielczarek:2016xql} the Spin-Field Correspondence has been studied focusing on the example of the Heisenberg XXX model in a constant magnetic field. It has been shown that such model is dual to a non-relativistic scalar field theory, characterised by quadratic dispersion relation. Due to the explicit breaking of the Lorentz symmetry, the recovered field theory is not relevant from the perspective of the physics of the fundamental interactions. Here, by generalising the Heisenberg XXX model to the Heisenberg XYZ model we are searching for a spin system which is dual (in the sense of the SFC) to the relativistic Klein-Gordon scalar field theory.

The XYZ model generalises the discrete Heisenberg model such that couplings between spin components may differ, namely 
\begin{equation}
H =- \sum_{i,j} \vec{S}'_i \cdot  \vec{S}_j - \mu \sum_i \vec{B} \cdot  \vec{S}_i,
\label{Heisenberg}
\end{equation}
where $\vec{S}'=\{J_xS_x,J_yS_y,J_zS_z\}$ and $J_x,\,J_y,\,J_z$ are real directional coupling constants, while $\mu$ are the magnetic coupling constant. The first sum is performed over the nearest neighbours. Here, the three dimensional case is considered, such that $\Sigma=\mathbb{R}^3$. Fundamentally, we consider our theory to be described by the discrete Hamiltonian (\ref{Heisenberg}). However, we assume here that the lattice spacing is below accuracy of any available apparatus, such that the continuous approximation of the system is valid. In practice, while considering the emergent field as a fundamental field, the lattice spacing should be below the current threshold which is roughly $1/(13\ \text{TeV}) \sim 10^{-19} $m. Of course, this holds under the assumption that the scalar field we are considering can be experimentally probed.  

While passing to the continuous approximation, the $i$ indices are replaced (in 1D) by the $x$ coordinate, 
such that for the neighbouring spins
\begin{equation}
\vec{S}'_i\cdot  \vec{S}_{i+1}\rightarrow  \vec{S}'(x)\cdot  \vec{S}(x+\epsilon)= 
\vec{S}'(x)\cdot  \vec{S}(x)+\epsilon \vec{S}'(x) \cdot \left.\frac{d \vec{S}}{dx} \right|_{\epsilon=0}+
\frac{1}{2} \epsilon^2 \vec{S}'(x) \cdot \left.\frac{d^2 \vec{S}}{dx^2} \right|_{\epsilon=0}+\mathcal{O}(\epsilon^3),
\end{equation}
with some infinitesimal parameter $\epsilon$. Within the case of the XXX Heisenberg model the second term in the Taylor expansion is vanishing, since $0=\frac{d (\vec{S} \cdot \vec{S})}{dx} = 2 \vec{S} \cdot \frac{d \vec{S}}{dx}$. Here, reshuffling the symmetrised contributions $\vec{S}'_i\cdot  \vec{S}_{i+1}$  and $\vec{S}_i\cdot  \vec{S}'_{i+1}$, the first order derivative leads to the total derivative $\frac{d (\vec{S}' \cdot \vec{S})}{dx}$, which contributes as a constant factor to the Hamiltonian, and is then subtracted. Notice that the first term in the Taylor expansion corresponds to the self-interactions of spins ($\vec{S}'_i \cdot  \vec{S}_i$), which is not present in the discrete case and, therefore, has to be subtracted from the continuos Hamiltonian. In the continuous limit, the sum is replaced by the integral $\sum_i  \rightarrow \frac{1}{\epsilon} \int dx $. As a consequence, the couplings $J_i$ have to be replaced by its densitised counterparts, i.e. $J_i \rightarrow \frac{\tilde{J}_i}{\epsilon}$.  Performing the $\epsilon \rightarrow 0$ limit, terms $\mathcal{O}(\epsilon^3)/\epsilon^2$ are then suppressed. Finally, only the second order derivative and the interaction term contribute to the continuous Hamiltonian.

The same conclusion can be drawn un the 3D case, for which the continuous version of the 
Hamiltonian (\ref{Heisenberg}) can be written as 
\begin{equation}
H = \int d^3x \, \mathcal{H} =  \int d^3x \left[ (\nabla \vec{S}' )\cdot(\nabla \vec{S} ) - \tilde{\mu}\, \vec{B} \cdot  \vec{S}  \right],
\label{HeisenbergCont}
\end{equation}
where $\vec{S}'=\{\tilde{J}_xS_x,\tilde{J}_yS_y,\tilde{J}_zS_z\}$ and 
$\tilde{J}_x,\,\tilde{J}_y,\,\tilde{J}_z$ are directional coupling constants 
for the continuous model, while $\tilde{\mu}$ stands for the continuous equivalent
of the magnetic coupling constant.

By applying the relations (\ref{Sx})-(\ref{Sz}) expanded up to the desired order in $\pi_\varphi/R_2$ and $\varphi/R_1$, 
the Hamiltonian density (\ref{HeisenbergCont}) can be recast as $\mathcal{H} = \mathcal{H}_0 +\mathcal{H}_{int}$, 
where the free field part is
\begin{equation}
\mathcal{H}_0=-SM-\sqrt{S}\bigg(\tilde{\mu}B_y\sqrt{M}\varphi+\frac{\tilde{\mu}B_z}{\sqrt{M}}\pi_\varphi\bigg)
+ \frac{\pi_\varphi^2}{2}+\frac{1}{2}(\nabla  \varphi)^2+\frac{1}{2}M^2 \varphi^2 +S\frac{\tilde{J}_z}{2M}(\nabla  \pi_\varphi)^2,
\label{HamField}
\end{equation}
and the interaction Hamiltonian, expanded up to the quartic order, reads
\begin{equation}
\begin{split}
\mathcal{H}_{int}&=\frac{1}{SM}\bigg[\sqrt{S}\bigg(\frac{1}{2}\tilde{\mu}B_y\sqrt{M}\varphi\pi_\varphi^2
+\frac{1}{6}\tilde{\mu}B_yM^{\frac{3}{2}}\varphi^3+\frac{\tilde{\mu}B_z}{6\sqrt{M}}\pi_\varphi^3\bigg)
\\&
-\frac{1}{24}\pi_\varphi^4-\frac{M^4}{24} \varphi^4-\frac{M^2}{4} \varphi^2\pi_\varphi^2
+\big(\tilde{J}_xSM-1\big)\big(2\varphi\pi_\varphi(\nabla\varphi)(\nabla\pi_\varphi)+\varphi^2(\nabla\varphi)^2\big)
\\&
+\frac{(\tilde{J}_x-\tilde{J}_z)S}{M}\pi_\varphi^2(\nabla \pi_\varphi)^2-\pi_\varphi^2(\nabla \varphi)^2
\bigg]+\mathcal{O}(6).
\end{split}
\label{HamIntKrak}
\end{equation}
In the latter equation we used the relations $R_1=\sqrt{\frac{S}{M}}$, $\tilde{J}_y=\frac{1}{SM}$ and $\tilde{\mu}B_x=M$ in order to reproduce the terms that describe the dynamics of the Klein-Gordon field. The case studied in Ref.~\cite{Mielczarek:2016xql} is recovered by fixing $B_y=0=B_z$ and by choosing all the spin couplings to be equal, i.e. $\tilde{J}_x=\tilde{J}_y=\tilde{J}_z$.

The expansion discussed above has been performed according to the powers of the 
field variables $\pi_\varphi$ and $\varphi$. However, after the coefficients of the terms contributing to 
of the Klein-Gordon field Hamiltonian are fixed in agreement with the standard case the powers of the 
scalar field variables in the perturbative expansion became correlated with the increasing negative 
powers of the total spin $S$. Higher the value of $S$, lower the curvature of the phase space is 
and in consequence smaller the contributions from the higher powers of the field variables are. 
Therefore, the free field case (quadratic Hamiltonian), is recovered in the large value of spin $S$ limit. 
This can be interpreted as the classical limit of the spin variables.

\subsection{Relativistic scalar field theory}

Let us now analyse the free Hamiltonian (\ref{HamField}) from the perspective of reconstructing the relativistic theory. By fixing the vector $\vec{B}$  to be oriented in the $x$ direction, the linear contributions to the Hamiltonian are vanishing. The remaining Lorentz symmetry violating part is the term $S\frac{\tilde{J}_z}{M}(\nabla  \pi_\varphi)^2$. This can be eliminated by suppressing the interactions between the $z$ components of the spin, which can be introduced by setting the coupling constant $\tilde{J}_z=0$.

In general, the contribution from the $z$ component can be parametrised by the dimensionless anisotropy parameter $\Delta$, such that $\tilde{J}_z = \Delta \tilde{J}$  and $\tilde{J}_x=\tilde{J}_y=\tilde{J}$, which corresponds to the so-called XXZ model. By assuming that $B_y=0=B_z$, the Hamiltonian density reduces to
\begin{equation}
\mathcal{H}_0=-SM+ \frac{\pi_\varphi^2}{2}+\frac{1}{2}(\nabla  \varphi)^2+\frac{1}{2}M^2 \varphi^2 
+\frac{\Delta}{2M^2}(\nabla  \pi_\varphi)^2,
\label{HamField2}
\end{equation}
which up to the new $\frac{\Delta}{2M^2}(\nabla  \pi_\varphi)^2$ term corresponds to the Hamiltonian of the Klein-Gordon field. The constant contribution $-SM$ leads to an energy shift which has no relevance in the classical theory.  

The choice of the $\vec{B}$ field to be aligned along the $x$ axis (such that  $B_y=0=B_z$), which leads to the Hamiltonian (\ref{HamField2}), is of course not restrictive, being symmetric under the choice of the orientation of the vector $\vec{B}$. The relevant point in stead is that the direction selected by $\vec{B}$ defines the equilibrium state with respect to which the $\varphi$ and $\pi_\varphi$ variables are defined. In other words, switching on the $\vec{B}$ vector amounts to induce to the spin chain system a spontaneous symmetry breaking (SSB). This in turn corresponds to select one of the available (degenerate) vacua of the Heisenberg XXX model.
We have chosen the $\vec{B}$ vector in such a way to fit to the definitions of the field variables introduced earlier. However, in general, the direction of the $\vec{B}$ field can be arbitrary, and the choice of a particular direction may result as a consequence of the SSB. Then, having the given direction pointed by $\vec{B}$, recovering of the Klein-Gordon field requires that the anisotropy of the spin model is introduced in the direction normal to $\vec{B}$. In the example considered here the $\vec{B}$ vector is directed along $x$ axis, while anisotropy is present in the $z$ directions. In the interesting case, when $\Delta \rightarrow 0$ and the interactions between spins in the $z$ direction are vanishing, the relativistic symmetry is recovered at the linear order. We may then argue that for $\Delta\neq1$ the SSB triggers a departure from the Lorentz symmetry. Conversely, the case in which $\Delta=0$ corresponds to setting an effective scenario in which spin interactions are confined to bidimensional layers that are localized on $y$-$z$ planes. Spin interactions are not affected then by the presence of the $\vec{B}$ field, aligned along the orthogonal $x$ direction.

\begin{figure}[ht!]
\centering
\includegraphics[width=8cm, angle = 0]{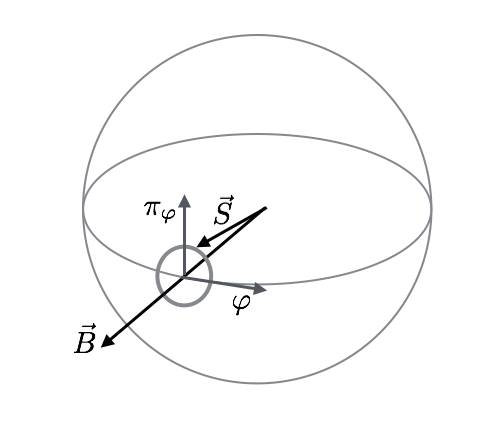}
\caption{Precession of the spin vector $\vec{S}$ around the magnetic field vector $\vec{B}$.}
\label{Fig2}
\end{figure}

In the linear regime, at each space point the vector $\vec{S}$ is precessing around the $\vec{B}$ vector, as shown in Fig. \ref{Fig2}. The top of the $\vec{S}$ vector encircles an ellipse at the phase space. In the linear regime, this is just a circle on the $(\varphi,\pi_{\varphi})$ phase space, as expected for the case of the free scalar field.

It is worth noticing that for $\Delta = 1$ and $S\rightarrow \infty$ the theory is invariant under the reciprocity transformation:
\begin{align}
\varphi &\rightarrow \frac{\pi_{\varphi}}{M},   \\
\pi_{\varphi} &\rightarrow - \varphi M. 
\end{align}
Such type of symmetry was postulated by Max Born in the first half of the last century \cite{Born49}. In our case, both the free Hamiltonian density 
\begin{equation}
\mathcal{H}_0 \left[\varphi \rightarrow \frac{\pi_{\varphi}}{M},\pi_{\varphi} \rightarrow - \varphi M   \right] 
=   \mathcal{H}_0 \left[\varphi,\pi_{\varphi}\right], 
\end{equation}
and the Poisson bracket 
\begin{equation}
\left\{\varphi \rightarrow \frac{\pi_{\varphi}}{M},\pi_{\varphi} \rightarrow - \varphi M   \right\} 
=  \{\varphi, \pi_{\varphi} \} \end{equation}
are invariant with respect to the duality-symmetry among field value and momentum. Here one can see that the Born reciprocity can be a symmetry of theory for a certain value of a parameter of the theory ($\Delta =1$) and the symmetry transforms into relativistic symmetry in a certain limit ($\Delta \rightarrow 0$).  

Neglecting the higher nonlinear terms, the equations of motion corresponding to $\mathcal{H}_0$ are
\begin{eqnarray}
\dot{\varphi} &=& \pi_{\varphi} -\frac{\Delta}{M^2} \nabla^2 \pi_{\varphi}, \label{HE1} \\
\dot{\pi}_{\varphi} &=& - M^2 \varphi + \nabla^2 \varphi, \label{HE2}
\end{eqnarray}
which lead to the modified version of the Klein-Gordon equation
\begin{equation}
\ddot{\varphi}-\left(1+\Delta\right) \nabla^2\varphi+M^2\varphi+\frac{\Delta}{M^2} \nabla^4\varphi =0.
\label{KGMod}
\end{equation}
The relativistic case is then recovered in the $\Delta\rightarrow 0$ limit.

\subsection{Dispersion relation}

Performing the Fourier transform
\begin{equation}
\varphi(t,\vec{x} )= \int \frac{d^3 p\, dE}{(2\pi)^4} \varphi(\omega,\vec{k} )e^{i(\vec{p} \cdot \vec{x}-E t)},
\end{equation}
the equation (\ref{KGMod}) immediately entails the following dispersion relation
\begin{equation}
E^2 = \left(1+\Delta\right)p^2+M^2 +\Delta \frac{p^4}{M^2} =  (p^2+M^2)\left(1+\Delta \frac{p^2}{M^2} \right).
\label{Dispersion}
\end{equation}
Three special cases for the dispersion relations are worth to be considered: 
\begin{itemize}
\item $\Delta=1$, for which $E = M +\frac{p^2}{M}$. This case corresponds to the non-relativistic model that satisfies the Born reciprocity symmetry for $S\rightarrow 0$. At the level of the spin system, this choice amounts to considering the XXX Heisenberg model. 
\item $\Delta=0$, for which $E^2=M^2+p^2$. This case amounts to the standard relativistic theory. 
\item $\Delta=-1$, for which $E^2=M^2-\frac{p^4}{M^2}$. This case corresponds to an unstable fixed point in the spin system. The instability can be easily concluded from the form of the dispersion relation.  
\end{itemize}

Furthermore, the dispersion relation (\ref{Dispersion}) rewrites, squared and expanded, as follows
\begin{equation}
E =\sqrt{p^2+M^2} \left(1+\Delta \frac{p^2}{2M^2}+\mathcal{O}(\Delta^2) \right).
\end{equation}
From the latter equation we can easily derive the group velocity
\begin{equation}
v_{gr} := \frac{\partial E}{\partial p} = \frac{p}{E} \left[1+\Delta\left(1+ \frac{p^2}{M^2} \right)\right].
\end{equation}
Consequently, we find the relation 
\begin{equation}
v_{gr} v_{ph}= 1+\Delta\left(1+ \frac{p^2}{M^2} \right),
\end{equation}
which might be both greater and smaller that than one, depending on the sign of $\Delta$.

\subsection{Interactions is case of the $XXZ$ model}

It is worth discussing in more details the spacial case of the vanishing anisotropy parameter $\Delta$. For this choice the relativistic Klein-Gordon theory is covered at the linear order.  
Structure of the interaction terms, which in the leading order contribute with 
$\mathcal{O}(1/S)$ deserve analysis. 

By choosing $\Delta=0$, together with $B_y=0=B_z$, the leading order interaction 
Hamiltonian (\ref{HamIntKrak}) reduces to:  
\begin{equation}
\begin{split}
\mathcal{H}_{int}&=\frac{1}{SM}\bigg[\!-\frac{1}{24}\pi_\varphi^4-\frac{M^4}{24} \varphi^4-\frac{M^2}{4} \varphi^2\pi_\varphi^2
+\frac{1}{M^2}\pi_\varphi^2(\nabla \pi_\varphi)^2-\pi_\varphi^2(\nabla \varphi)^2
\bigg]\!+\mathcal{O}\bigg(\frac{1}{S^2}\bigg).
\end{split}
\label{HamIntXY}
\end{equation}

As a consequence, the Hamilton equations for $\mathcal{H}=\mathcal{H}_0+\mathcal{H}_{int}$ read
\begin{equation}
\begin{split}
\dot{\varphi}&=\{\varphi,\mathcal{H}\}=\frac{1}{\cos(\pi_{\varphi}/\sqrt{SM})}
\bigg(\frac{\partial}{\partial\pi_{\varphi}}-\nabla\frac{\partial}{\partial(\nabla\pi_{\varphi})}\bigg)\mathcal{H}
\\
&=\pi_{\varphi}+\frac{1}{SM}\bigg(\frac{1}{3}\pi_{\varphi}^3-\frac{M^2}{2} \varphi^2\pi_\varphi
-\frac{2}{M^2}(\nabla \pi_\varphi)^2\pi_\varphi-\frac{2}{M^2}\big(\nabla^2 \pi_\varphi\big)\pi_\varphi^2-
2(\nabla \varphi)^2\pi_\varphi\bigg)
\\
&+\mathcal{O}\bigg(\frac{1}{S^2}\bigg)\,,
\end{split}
\label{DotPhi}
\end{equation}
\begin{equation}
\begin{split}
\dot{\pi}_{\varphi}&=\{\pi_{\varphi},\mathcal{H}\}=-\frac{1}{\cos(\pi_{\varphi}/\sqrt{SM})}
\bigg(\frac{\partial}{\partial\varphi}-\nabla\frac{\partial}{\partial(\nabla\varphi)}\bigg)\mathcal{H}
\\
&=-M^2\varphi+\nabla^2\varphi
+\frac{1}{SM}\bigg(\frac{M^4}{6}\varphi^3-4(\nabla\varphi)(\nabla\pi)\pi-\frac{3}{2}\big(\nabla^2\varphi\big)\pi^2\bigg)
+\mathcal{O}\bigg(\frac{1}{S^2}\bigg)\,,
\end{split}
\end{equation}
where $\mathcal{O}(1/S^2)$ contains terms of power in field or momenta variables being 5 or higher.\\

In order to derive the inverse Legendre transformation, we need the following expression
\begin{equation}
\begin{split}
\pi_{\varphi}^2
&=\dot{\varphi}^2-\frac{1}{SM}\bigg(\frac{2}{3}\dot{\varphi}^4-M^2\varphi^2\dot{\varphi}^2
-\frac{4}{M^2}(\nabla\dot{\varphi})^2\dot{\varphi}^2-\frac{4}{M^2}\big(\nabla^2\dot{\varphi})\dot{\varphi}^3
-4(\nabla \varphi)^2\dot{\varphi}^2\bigg)
\\
&+\mathcal{O}\bigg(\frac{1}{S^2}\bigg)\,.
\end{split}
\label{DotPhi}
\end{equation}
Finally, we get the Lagrangian density
\begin{equation}
\begin{split}
\mathcal{L}&=\pi_{\varphi}\dot{\varphi}-\mathcal{H}
=SM+\frac{\dot{\varphi}^2}{2}-\frac{1}{2}(\nabla  \varphi)^2-\frac{1}{2}M^2\varphi^2
\\
&+\frac{1}{SM}\bigg(\frac{1}{24}\dot{\varphi}^4+\frac{M^4}{24}\varphi^4+\frac{M^2}{4}\varphi^2\dot{\varphi}^2
-\frac{1}{M^2}(\nabla\dot{\varphi})^2\dot{\varphi}^2+(\nabla \varphi)^2\dot{\varphi}^2
\bigg)+\mathcal{O}\bigg(\frac{1}{S^2}\bigg)\,.
\end{split}
\label{Lagrangian}
\end{equation}

This interaction Lagrangian can not be cast as the product of the two Casimir operators of the Poincar\'e algebra, thus there is no manifest symmetry in it. This situation is anyway 
reminiscent of the $\lambda \varphi^4$ interaction term for a real scalar scalar field 
theory that enjoys Poincar\'e symmetries. It is worth noticing though that contrary to the case of the $\lambda \varphi^4$ theory the Lagrangian density \eqref{Lagrangian} is not renormalisable in the classical sense, because of the presence at $\mathcal{O}(1/S)$ of irrelevant operators \cite{Peskin:1995ev}. So counter-terms to dimensional operators at $\mathcal{O}(1/S)$ will require the inclusion of irrelevant operators at $\mathcal{O}(1/S^2)$ and so forth.  This suggests that the (fundamental) theory to be considered at the quantum level is the XXZ Heisenberg model, the continuos scalar field theory only arising as an effective model. This remark might have a profound significance when extending the equivalence to the other matter fields and to gravity.

\section{$\Delta$-deformed or $\Delta$-broken symmetries?} \label{Sec_Deformations}

In this section we address the fate of the space-time symmetries for the effective (continuos) scalar field theory under consideration. Within the limit $\Delta \ll1$, which amounts to an orientation of the angular momentum along the plane orthogonal to the z-axis, we can ask ourselves whether the emergent space-time symmetries are still described by a Lie-algebra, or we should rather contemplate a breakdown of the Lorentz symmetries, or the emergence of a new symmetry described in terms of an Hopf algebra, representing the deformation of the symmetry manifest in the free Hamiltonian, which we considered  at zeroth order in $\Delta$.  We address these questions resorting to the constructions of the Hopf algebras --- for a tutorial see e.g. Refs.~\cite{quantum_groups1,quantum_groups2}. This is a mathematical structure that we are forced to deal with if we want to extend the Lie algebra like structure to bi-algebras or generic tensor products of copies of the same algebra. In this perspective, we can claim to have still a (co-commutative) Lie algebra only if we are still able to find a trivial (co-commutative) Hopf algebra structure for the `deformed' (or `broken') space-time symmetries.

\subsection{Hopf algebra?}

We start our analysis by first recovering the algebra of the space-time transformations. This can be achieved by casting the generators as functionals of the field operators of the theory. We can consistently follow three different strategies, and every time reach the same result. Specifically, we can: \\ \\

{\it i)} cast the theory as a second-order derivatives one, and apply the Noether theorem to the Lagrangian $\mathcal{L}=\mathcal{L}[\varphi, \partial_\mu\varphi, \partial_\mu\partial_\nu\varphi ]$; \\

{\it ii)} define a tensorial star-product between the fields that allows to cast the Lagrangian as
\begin{equation}
\mathcal{L}= \frac{1}{2} \partial_\mu \varphi \, \star^{\mu\nu} \partial_\nu \varphi -\frac{1}{2} M^2 \varphi^2\,.
\end{equation}
The star-product must be of the type
\begin{equation}
\star^{\mu\nu}={\rm diag}\left(\left(1-\frac{\Delta }{M^2} \overleftarrow{\nabla} \cdot \overrightarrow{\nabla} \right), -1, -1, -1\right)\,,
\end{equation}
as arises from direct inspection of the Lagrangian, and from a choice of arrangement of the fields derivatives;\\

{\it iii)} in an equivalent way, define
\begin{equation}
\star^{\mu\nu}={\rm diag}\left( 1, -\left(1+\frac{\Delta }{M^2} \overleftarrow{\nabla} \cdot \overrightarrow{\nabla} \right), -\left(1+\frac{\Delta }{M^2} \overleftarrow{\nabla} \cdot \overrightarrow{\nabla} \right), -\left(1+\frac{\Delta }{M^2} \overleftarrow{\nabla} \cdot \overrightarrow{\nabla} \right)\right)\,,
\end{equation}
leading to the same dispersion relation as in the case {\it ii)}.\\

Moving along the path described in {\it iii)}, we can express the generators of the algebra as functional of the phase-space variables that have the form
\begin{eqnarray}
&& \mathcal{P}_0=\mathcal{H}_0=\dot{\varphi} \, \pi_\varphi - \mathcal{L}\,,
\qquad \mathcal{P}_i=\pi_\varphi \star_{ij} \partial^j \varphi\,, \qquad \mathcal{J}_{\mu \nu} = x_\mu \mathcal{P}_\nu- x_\nu \mathcal{P}_\mu\,.
\end{eqnarray}

It is remarkable that the generators of the space-time transformation $X=\{\mathcal{P}_{\mu}\,,\mathcal{J}_{\mu \nu}\}$ fulfill up to the first order in $\Delta$ the Poisson brackets
\begin{eqnarray} \label{comrel}
&&\{ \mathcal{P}_{\mu}, \mathcal{P}_{\nu}\}=0\,, \nonumber\\
&& \{ \mathcal{P}_\mu, \mathcal{J}_{\nu \rho} \}= \imath (\eta_{\mu \rho} \mathcal{P}_\nu - \eta_{\mu \nu} \mathcal{P}_\rho)\,, \nonumber\\
&& \{ \mathcal{J}_{\mu \nu}, \mathcal{J}_{\sigma \rho} \}= \imath (\eta_{\mu \rho} \mathcal{J}_{\nu \sigma} - \eta_{\nu \rho} \mathcal{J}_{\mu \sigma} +  \eta_{\nu \sigma} \mathcal{J}_{\mu \rho} -  \eta_{\mu \sigma} \mathcal{J}_{\nu \rho} )\,,
\label{PoincareAlgebra}
\end{eqnarray}
which are proper of the Poincar\'e algebra. Nonetheless, the algebra has non-trivial co-product, with space-time 
coordinates dependence, resulting in the emergence at $\mathcal{O}(\Delta)$ of a non-trivial bi-algebroid.

We can now ask ourselves whether this bi-algebra is a Hopf algebra. The co-product, which is a map between the elements of the algebra and the elements of the tensor product of the two copies of the algebra, as well as the other applications that are involved in the enunciation of the Hopf algebra axioms \cite{quantum_groups1,quantum_groups2}, can be easily calculated when generators are represented as derivatives of space-time coordinates. In this case the representation of the algebra reads
\begin{eqnarray}
&& \mathcal{P}_0= \partial_0\,, \qquad \mathcal{P}_i= \left(1-\frac{\Delta }{M^2} \overrightarrow{\nabla}^2 \right) \, \partial_i , \qquad \mathcal{J}_{\mu \nu} = x_\mu \mathcal{P}_\nu- x_\nu \mathcal{P}_\mu\,,
\end{eqnarray}
which makes straightforward to recover respectively the co-product $\Delta(X)$, the antipode $S(X)$ and the co-unit $\epsilon(X)$ of the bi-algebra:
\begin{align}
\Delta(\mathcal{P}_0)=&\ \mathcal{P}_0 \otimes \mathds{1} + \mathds{1} \otimes \mathcal{P}_0,
\\
\Delta(\mathcal{P}_i)=&\ \mathcal{P}_i \otimes \mathds{1} + \mathds{1} \otimes \mathcal{P}_i
-\sum_k \frac{\Delta}{M^2}
\big(
2\,\partial_i\partial_k\otimes\partial_k+2\,\partial_k\otimes\partial_k\partial_i
+\partial_i\otimes\partial_k\partial_k+\partial_k\partial_k\otimes\partial_i
\big),
\\
\begin{split}
\Delta(\mathcal{J}_{ij})=&\ \mathcal{J}_{ij} \otimes \mathds{1} + \mathds{1} \otimes \mathcal{J}_{ij}
-\sum_k \frac{\Delta}{M^2}
\Big(
2\big(x_i\partial_j-x_j\partial_i\big)\partial_k\otimes\partial_k+2\,\partial_k\otimes\big(x_i\partial_j-x_j\partial_i\big)\partial_k
\\
&+\big(x_i\partial_j-x_j\partial_i\big)\otimes\partial_k\partial_k+\partial_k\partial_k\otimes\big(x_i\partial_j-x_j\partial_i\big)
\Big),
\end{split}
\\
\begin{split}
\Delta(\mathcal{J}_{0i})=&\ \mathcal{J}_{0i} \otimes \mathds{1} + \mathds{1} \otimes \mathcal{J}_{0i}
-\sum_k \frac{\Delta}{M^2}
\big(
2\,x_0\partial_i\partial_k\otimes\partial_k+2\,\partial_k\otimes x_0\partial_i\partial_k
\\
&+x_0\partial_i\otimes\partial_k\partial_k+\partial_k\partial_k\otimes x_0\partial_i
\big),
\end{split}
\end{align}

\begin{align}
S(\mathcal{P}_\mu)=-\mathcal{P}_\mu\,,
\qquad
S(\mathcal{J}_{\mu\nu})=-\mathcal{J}_{\mu\nu}\,,
\qquad
\epsilon(\mathcal{P}_\mu)=0\,,
\qquad
\epsilon(\mathcal{J}_{\mu \nu})=0\,.
\end{align}

To conclude we are dealing with a Hopf algebra, it is necessary to check whether on the algebra the co-product map acts as a morphism. A straightforward but lengthy calculation shows that:  

\begin{equation}
\Delta[\mathcal{P}_\mu, \mathcal{P}_\nu ]=[\Delta(\mathcal{P}_\mu), \Delta(\mathcal{P}_\nu)]=0\,,
\end{equation}
\begin{equation}
\Delta[\mathcal{P}_0, \mathcal{J}_{ij}]=[\Delta(\mathcal{P}_0), \Delta( \mathcal{J}_{ij})]=0\,,
\end{equation}
\begin{equation}
\begin{split}
\Delta[\mathcal{P}_0, \mathcal{J}_{0i}]
=&\,-\Delta(\mathcal{P}_i)
\\
=-[\Delta(\mathcal{P}_0), \Delta( \mathcal{J}_{0i})]
=&\,-\mathcal{P}_i \otimes \mathds{1}-\mathds{1} \otimes \mathcal{P}_i
+\sum_l\frac{\Delta}{M^2}
\big(
2\,\mathcal{P}_i\mathcal{P}_l\otimes\mathcal{P}_l+2\,\mathcal{P}_l\otimes\mathcal{P}_l\mathcal{P}_i
\\
&\,+\mathcal{P}_i\otimes\mathcal{P}_l\mathcal{P}_l
+\mathcal{P}_l\mathcal{P}_l\otimes\mathcal{P}_i
\big)+\mathcal{O}\bigg(\frac{\Delta^2}{M^4}\bigg)\,,
\end{split}
\end{equation}
\begin{equation}
\begin{split}
\Delta[\mathcal{P}_i, \mathcal{J}_{jk}]
=&\,-\eta_{ij}\Delta(\mathcal{P}_k)+\eta_{ik}\Delta(\mathcal{P}_j)
\\
=-[\Delta(\mathcal{P}_i), \Delta( \mathcal{J}_{jk})]
=&\,-\big(\eta_{ij}\mathcal{P}_k-\eta_{ik}\mathcal{P}_j\big)\otimes\mathds{1}
-\mathds{1}\otimes\big(\eta_{ij}\mathcal{P}_k-\eta_{ik}\mathcal{P}_j\big)
\\
&\,+\sum_l\frac{\Delta}{M^2}\Big(
2\big(\eta_{ij}\mathcal{P}_k-\eta_{ik}\mathcal{P}_j\big)\mathcal{P}_l\otimes\mathcal{P}_l
+2\,\mathcal{P}_l\otimes\big(\eta_{ij}\mathcal{P}_k-\eta_{ik}\mathcal{P}_j\big)\mathcal{P}_l
\\
&\,+\big(\eta_{ij}\mathcal{P}_k-\eta_{ik}\mathcal{P}_j\big)\otimes\mathcal{P}_l\mathcal{P}_l
+\,\mathcal{P}_l\mathcal{P}_l\otimes\big(\eta_{ij}\mathcal{P}_k-\eta_{ik}\mathcal{P}_j\big)
\Big)+\mathcal{O}\bigg(\frac{\Delta^2}{M^4}\bigg)\,,
\end{split}
\end{equation}
\begin{equation}
\begin{split}
\Delta[\mathcal{P}_i,\mathcal{J}_{0j}]
=&\ \eta_{ij}\Delta(\mathcal{P}_0)
\\
\neq-[\Delta(\mathcal{P}_i),\Delta( \mathcal{J}_{0j})]
=&\ \eta_{ij}\mathcal{P}_0\otimes\mathds{1}+\mathds{1}\otimes\eta_{ij}\mathcal{P}_0
-\sum_l\frac{\Delta}{M^2}\Big(
2\,\mathcal{P}_0\mathcal{P}_i\mathcal{P}_j\otimes\mathds{1}+\mathds{1}\otimes2\,\mathcal{P}_0\mathcal{P}_i\mathcal{P}_j
\\
&\,+2\,\mathcal{P}_0\mathcal{P}_i\otimes\mathcal{P}_j+2\,\mathcal{P}_j\otimes\mathcal{P}_i\mathcal{P}_0
+2\,\mathcal{P}_0\mathcal{P}_j\otimes\mathcal{P}_i+2\,\mathcal{P}_i\otimes\mathcal{P}_0\mathcal{P}_j
\\
&\,+2\,\mathcal{P}_i\mathcal{P}_j\otimes\mathcal{P}_0+2\,\mathcal{P}_0\otimes\mathcal{P}_i\mathcal{P}_j
+\eta_{ij}\big(
2\,\mathcal{P}_0\mathcal{P}_l\otimes\mathcal{P}_l+2\,\mathcal{P}_l\otimes\mathcal{P}_0\mathcal{P}_l
\\
&\,+\mathcal{P}_0\otimes\mathcal{P}_l\mathcal{P}_l+\mathcal{P}_l\mathcal{P}_l\otimes\mathcal{P}_0
\big)
\Big)+\mathcal{O}\bigg(\frac{\Delta^2}{M^4}\bigg)\,,
\end{split}
\end{equation}
\begin{equation}
\begin{split}
\Delta[\mathcal{J}_{0i}, \mathcal{J}_{0j}]
=&\,-\Delta(\mathcal{J}_{ij})
\\
\neq[\Delta(\mathcal{J}_{0i}), \Delta( \mathcal{J}_{0j})]
=&\,-\mathcal{J}_{ij}\otimes\mathds{1}-\mathds{1}\otimes\mathcal{J}_{ij}
\end{split}
\end{equation}
\begin{equation}
\begin{split}
\Delta[\mathcal{J}_{ij}, \mathcal{J}_{0k}]
=&\ \eta_{ik}\Delta(\mathcal{J}_{j0})-\eta_{jk}\Delta(\mathcal{J}_{i0})
\\
\neq[\Delta(\mathcal{J}_{ij}), \Delta( \mathcal{J}_{0k})]
=&\ \big(\eta_{ik}\mathcal{J}_{j0}-\eta_{jk}\mathcal{J}_{i0}\big)\otimes\mathds{1}
+\mathds{1}\otimes\big(\eta_{ik}\mathcal{J}_{j0}-\eta_{jk}\mathcal{J}_{i0}\big)
\\
&\,-\sum_l\frac{\Delta}{M^2}\bigg(
2\big(\eta_{ik}\mathcal{J}_{j0}-\eta_{jk}\mathcal{J}_{i0}\big)\mathcal{P}_l\otimes\mathcal{P}_l
+2\,\mathcal{P}_l\otimes\big(\eta_{ik}\mathcal{J}_{j0}-\eta_{jk}\mathcal{J}_{i0}\big)\mathcal{P}_l
\\
&\,+\big(\eta_{ik}\mathcal{J}_{j0}-\eta_{jk}\mathcal{J}_{i0}\big)\otimes\mathcal{P}_l\mathcal{P}_l
+\mathcal{P}_l\mathcal{P}_l\otimes\big(\eta_{ik}\mathcal{J}_{j0}-\eta_{jk}\mathcal{J}_{i0}\big)
\\
&\,-2\,\mathcal{P}_0\otimes\mathcal{J}_{ij}\mathcal{P}_k
-2\,\mathcal{J}_{ij}\mathcal{P}_k\otimes\mathcal{P}_0
-2\,\mathcal{P}_0\mathcal{P}_k\otimes2\,\mathcal{J}_{ij}
-2\,\mathcal{J}_{ij}\otimes\mathcal{P}_0\mathcal{P}_k
\bigg)\,,
\end{split}
\end{equation}
\begin{equation}
\begin{split}
\Delta[\mathcal{J}_{ij}, \mathcal{J}_{kl}]
=&\ \eta_{il}\Delta(\mathcal{J}_{jk})-\eta_{jl}\Delta(\mathcal{J}_{ik})+\eta_{jk}\Delta(\mathcal{J}_{il})-\eta_{ik}\Delta(\mathcal{J}_{jl})
\\
\neq[\Delta(\mathcal{J}_{ij}), \Delta( \mathcal{J}_{kl})]
=&\ [\mathcal{J}_{ij}, \mathcal{J}_{kl}]\otimes\mathds{1}
+\mathds{1}\otimes[\mathcal{J}_{ij}, \mathcal{J}_{kl}]
\\
&\,-\sum_m\frac{\Delta}{M^2}\big(
4[\mathcal{J}_{ij}, \mathcal{J}_{kl}]\mathcal{P}_m\otimes\mathcal{P}_m
+4\,\mathcal{P}_m\otimes[\mathcal{J}_{ij}, \mathcal{J}_{kl}]\mathcal{P}_m
\\
&\,+2[\mathcal{J}_{ij}, \mathcal{J}_{kl}]\otimes\mathcal{P}_m\mathcal{P}_m
+2\,\mathcal{P}_m\mathcal{P}_m\otimes[\mathcal{J}_{ij}, \mathcal{J}_{kl}]\big)\,.
\end{split}
\end{equation}

This is enough to infer that the bi-algebroid we recovered is not a Hopf algebra. As consequence, although the commutation relations in \eqref{comrel} would have let us think so, at $\mathcal{O}(\Delta)$ there is not even a Lie algebra. This is relevant to the construction of the Fock space of the theory, and to discuss the statistics of multi-particles states. Consistently with this result, in the next subsection we show that is possible to recover a basis of generators for the space-time transformation in which the breakdown of the Lorentz (Poincar\'e) symmetry does not show up only at the level of the co-algebra, but is manifest also in the algebraic sector. 

The issue of the characterization of space-time transformation symmetries received much attention in the last decade, especially from the perspective of studies devoted first to doubly special relativity and then to relative locality. Within this framework, we emphasize that a description of the deformed symmetries would necessarily require also a characterization of the conserved quantities associated in a fashion that must independent
of the choice of algebraic basis --- see e.g. the studies of deformed symmetries in Refs.~\cite{Agostini:2006nc,Arzano:2007gr,Arzano:2007ef,AmelinoCamelia:2007uy,AmelinoCamelia:2007vj,AmelinoCamelia:2007wk,AmelinoCamelia:2007rn,Marciano:2008tva,Marciano:2010jm}. Possible phenomenological applications have been also considered  --- see e.g. Refs~\cite{AmelinoCamelia:2007zzb, Marciano:2010gq, AmelinoCamelia:2010zf, AmelinoCamelia:2012it}. Finally, we comment on the fact the emergence of deformed symmetries is by some authors conjectured (and in some cases recovered) to be a feature of effective theories derived from more fundamental non-perturbative models of quantum geometry \cite{MarMod,NClimLQG,BRAM,BraMaRo}. It is then interesting to notice the similarity that is present, at this level, with the emergence of continuos (scalar) field theories from a ``fundamental'' spin-chain theory. This latter, as we emphasized in the introduction, is very much reminiscent of the string-nets attempt in condensed matter to reconstruct gauge field structures \cite{LW,WenBook,Gu:2010na}. Furthermore, this very same lattice structure, colored by irreducible representations, naturally suggests that the SFC could be extended so to account for loop quantum gravity \cite{carlo,thomas,alex}, as we will argue better in Sec.~\ref{Sec_Polymer}.

\subsection{Algebraic sector from the modified dispersion relation}

An alternative method for deriving the algebraic sector of the space-time symmetries is to recover the algebra 
of the space-time transformations by analysing transformation properties of the modified dispersion relation 
(\ref{Dispersion}) given by
\bea
g^2(p) E^2 - p^2 = M^2\,,
\eea
with $g^2(p)=1/\left[1+\Delta\left(\frac{p^2}{M^2}\right)\right]$.  The dispersion relation should be invariant under 
generators of space-time transformations that show a $\Delta$-departure from the Lorentz transformations, and 
are such that $g^2(p') (E')^2 - (p')^2 = g^2(p) E^2 - p^2 = M^2$ is satisfied. Our aim is to derive the modified 
Lorentz transformations, in terms of the original basis $(E,p_i)$, which is schematically given by
\bea
\left(
  \begin{array}{c}
    E' \\
    p_x^\prime \\
    p_y^\prime \\
    p_z^\prime \\
  \end{array}
\right) = \tilde{\Lambda}  \left(
  \begin{array}{c}
    E \\
    p_x \\
    p_y \\
    p_z \\
  \end{array}
\right)\,,
\eea
with
\bea
\tilde{\Lambda} = \left(
                    \begin{array}{cccc}
                      g^{-1}(p) & 0 & 0 & 0 \\
                      0 & 1 & 0 & 0 \\
                      0 & 0 & 1 & 0 \\
                      0 & 0 & 0 & 1 \\
                    \end{array}
                  \right)
                  \left(
                    \begin{array}{cccc}
                      \gamma & -\gamma v & 0 & 0 \\
                      -\gamma v & \gamma & 0 & 0 \\
                      0 & 0 & 1 & 0 \\
                      0 & 0 & 0 & 1 \\
                    \end{array}
                  \right)
                  \left(
                    \begin{array}{cccc}
                      g(p) & 0 & 0 & 0 \\
                      0 & 1 & 0 & 0 \\
                      0 & 0 & 1 & 0 \\
                      0 & 0 & 0 & 1 \\
                    \end{array}
                  \right)
\eea

A little bit of algebra yields that the Lorentz like transformations are given by
\bea
E' &=& Q\gamma \left[\frac{E}{\sqrt{1+\Delta \left(\frac{p^2}{M^2}\right)}} - v p_x\right],\\
p_x^\prime &=& \gamma \left[p_x - \frac{vE}{\sqrt{1+\Delta \left(\frac{p^2}{M^2}\right)}} \right],\\
p_y^\prime &=& p_y,\\
p_z^\prime &=& p_z,
\eea
where
\bea
Q = \sqrt{1 + \frac{\Delta}{M^2} \left[\gamma^2 \(p_x - \frac{v E}{\sqrt{1+\(\frac{\Delta}{M^2}\)p^2}}\)^2 + p_y^2 + p_z^2\right]}.
\eea

Obviously, in the explicit calculation above, we only considered a boost along the $x$-direction. 
However, this can be easily generalised. The algebraic sector of the generators of the space-time
transformations can be derived by first defining $\Delta$-independent coordinates,
\bea
\pi^0 &=& g(p) E\,, \hspace{2cm} \pi^i = p^i\,,\\
\xi^0 &=& \frac{1}{g(p)}\,x^0\,, \hspace{2cm} \xi^i = x^i\,.
\eea
The new variables transform under usual special relativistic laws and thus $\pi^\mu \pi_\mu = g^2(p)E^2 - p^2 = M^2$ looks like a Casimir invariant. Since $\(\xi, \pi\)$ forms a usual canonical pair, we can choose a particular basis to write
\bea
\xi_\mu = i\frac{\partial}{\partial \pi^\mu}\,,
\eea
and we have the relations
\bea
[\pi_\mu , \pi_\nu] = 0 = [\xi^\mu, \xi^\nu]\,, \ \ \
[\xi^\mu , \pi_\nu] = \delta^\mu_\nu\,.
\eea

Having defined these auxiliary variables, we look to get the commutator of the generators of the algebra of space-time transformations in terms of the original ones. To this end, we start by applying the chain rule to get
\bea
\frac{\partial}{\partial \pi^0} &=& \frac{1}{g(p)}\frac{\partial}{\partial E}\,,\\
\frac{\partial}{\partial \pi^i} &=& \(\frac{\partial g}{\partial p^i}\) \(\frac{E}{g(p)}\)\frac{\partial}{\partial E}\,.
\eea
Therefore, the position operator, using $x_\mu = g(p) \xi_\mu$, is given by
\bea
x_0 &=& i\frac{\partial}{\partial E}\,,\\
x_i &=& i\(\frac{\partial}{\partial p^i} -  \(\frac{\partial g}{\partial p^i}\) \(\frac{E}{g(p)}\)\frac{\partial}{\partial E}\).
\eea

We can rewrite the position operator in a slightly more covariant manner as
\bea
x_\mu &=& i\(\frac{\partial}{\partial p^\mu} -  a_\mu E\frac{\partial}{\partial E}\)\,,
\eea
with $a_\mu := \delta^i_\mu \(\frac{\partial g}{\partial p^i}\) \(\frac{1}{g(p)}\)$.
The Lorentz like generators acquire now a $\Delta$-dependence
\bea
M_{\mu\nu} := p_\nu x_\mu  - p_\mu x_\nu = i\(p_\nu \frac{\partial}{\partial p^\mu} - p_\mu \frac{\partial}{\partial p^\nu} + \Theta_{\mu\nu} E \frac{\partial}{\partial E}\)\,,
\eea
with $\Theta_{\mu\nu} := p_\mu a_\nu - p_\nu a_\mu$.

The commutator between them remains the same as in the standard case
\bea
\left[M_{\mu\nu} , M_{\rho\lambda}\right] = i\(g_{\mu\rho}M_{\nu\lambda} - g_{\nu\rho}M_{\mu\lambda} - g_{\mu\lambda}M_{\nu\rho}  + g_{\nu\lambda}M_{\mu\rho}\)\,,
\eea
although the generators themselves are now $\Delta$-dependent. This confirms the results derived in the previous subsection --- see e.g. Eq.~\ref{PoincareAlgebra}. However, other commutators are necessarily $\Delta$-dependent
\bea
\left[M_{\mu\nu} , x_\rho \right] &=& i\(g_{\mu\rho}x_\nu - g_{\nu\rho}x_\mu - \Theta_{\mu\nu}x_\rho\)\,,\\
\left[M_{\mu\nu} , p_\rho \right] &=& i\(g_{\mu\rho}p_\nu - g_{\nu\rho}p_\mu + \Theta_{\mu\nu}p_\rho\)\,.
\eea

Finally, the Heisenberg algebra also necessarily develop a $\Delta$-dependence, giving rise to non-commutativity
\bea
\left[p_\mu, p_\nu\right] &=& 0\,,\\
\left[p_\mu, x_\nu\right] &=& i \(g_{\mu\nu} - a_\mu p_\nu\)\,,\\
\left[x_\mu, x_\nu\right] &=& i\(a_\mu x_\nu - a_\nu x_\mu\)\,.
\eea
The way we have evaluated the form of the algebra of space-time transformations, from the $\Delta$-dependent 
dispersion relation, goes along the lines of what is sometimes done in doubly-special relativity and non-commutative geometry theories. Worth noticing is that the obtained deformed version of the Heisenberg algebra 
shares  similarities with the case of $\kappa-$Minkowski space-time algebra \cite{Lukierski:1991pn,Lukierski:1993wxa}. 

\section{Relation with polymerisation} \label{Sec_Polymer}

The requirement of compactness for the field phase space, as the one implemented within SFC, may be considered as a guiding idea in the search for a quantum theory of the gravitational interaction. In this section, we contribute to analyse such a possibility by exploring the relation between the spherical phase space and the procedure of ``polymerisation'' related with the loop quantisation \cite{Ashtekar:2002sn}. 

In polymer quantum mechanics, the Hilbert space is such that the action of the $\hat{\pi}_{\varphi}$ operator is not well defined. Nonetheless, the action of  $\hat{U}_{\lambda}=\widehat{e^{i \lambda \pi_{\varphi}}}$ can be consistently implemented --- i.e. $\hat{U}_{\lambda}| \varphi \rangle = | \varphi+ \lambda \rangle$. The constant $\lambda$ is a new length scale, called scale of polymerisation. In 1D ($\Sigma=\mathbb{R}$), the action of the operator $\hat{U}_{\lambda}$ interpolates between the states defined at the set of points 
$\Gamma = \left\{ \varphi_i \right\} \subset \Sigma$, forming a graph. For equally distant points, we obtain $\varphi_i=i \lambda+\epsilon$, with different equivalent representations distinguished by the values of $\epsilon$. 

While the action of the operator $\hat{\pi}_{\varphi}$ is not well defined on the $\mathcal{H}_{\text{poly}}=$ span$\{|i \lambda+\epsilon \rangle\}$ Hilbert space, one can defined a polymerised version of the momentum operator, i.e. 
\begin{equation}  
\hat{\pi}_{\varphi} \rightarrow  \frac{ \hat{U}_{\lambda}-\hat{U}^{\dagger}_{\lambda}}{2 i \lambda}  =  
\widehat{\frac{\sin(\pi_{\varphi} \lambda)}{ \lambda}}, \label{polymerp}
\end{equation}
such that its action on $\mathcal{H}_{\text{poly}}$ is properly defined.   

The so-called polymerisation of momentum effectively introduces bending (periodification) of the 
phase space in the direction of $\pi_{\varphi}$, such that the classical phase space 
$\Gamma=\mathbb{R} \times \mathbb{R} \ni  (\varphi, \pi_{\varphi})$ is deformed to 
$\Gamma_{\text{poly}}=\mathbb{R} \times S^1$ \cite{Bojowald:2011jd}.

In what follows, we show that there is actually a connection between the polymerisation of the phase, 
which is related to the polymer quantisation. Let us discuss this relation focusing on the example of a 
harmonic oscillator. To facilitate comparison with standard literature, we begin by introducing a canonical 
transformation of the form
\bea
\pi_{\varphi} \rightarrow  -\varphi\, , \ \ \ \ 
\varphi \rightarrow \pi_\varphi\,.
\eea
Physically, this simply amounts to interchanging the labels of the phase space angles with the field and its conjugate momenta. In terms of this new canonical transformation, the spin components may be the recast as
\begin{align}
S_x:=&\  S \cos \left( \frac{\pi_{\varphi}}{R_1} \right) \cos \left( \frac{\varphi}{R_2} \right),
\label{SxPol} \\
S_y:=&\ S \sin \left( \frac{\pi_{\varphi}}{R_1} \right) \cos \left( \frac{\varphi}{R_2} \right),
\label{SyPol}\\
S_z:=&\ - S \sin \left( \frac{\varphi}{R_2} \right)\,.
\label{SzPol}
\end{align}
The corresponding symplectic form is given by
\bea
\omega = \cos\(\frac{\varphi}{R_2}\) \d\pi_\varphi \wedge \d\varphi\,.
\eea

For the case of a harmonic oscillator, the spin Hamiltonian can be written as
\begin{eqnarray}
 H = \alpha\left(1-\frac{S_x}{S}\right)\,. \label{HamHO}
\end{eqnarray}
Here $\alpha$ is a dimensional parameter introduced to give the correct units to the Hamiltonian. 
The fact that this is indeed the correct Hamiltonian for a harmonic oscillator can be easily verified 
by expanding the spin component in the small angle limit, as before
\bea
H = \alpha\frac{\pi_\varphi^2}{2R_1^2} + \alpha\frac{\varphi^2}{2R_2^2} + \mathcal{O}(4)\,.
\eea
The mass and the frequency can be identified respectively with
\bea
m=\frac{R_1^2}{\alpha} , \ \ \ \ \tilde{\omega} = \frac{\alpha}{S}\,,
\eea
where we have used $R_1R_2=S$. Thus, up to leading order we find the standard, classical harmonic oscillator Hamiltonian
\bea
H= \frac{\pi_\varphi^2}{2m} + \frac{m\tilde{\omega}^2\varphi^2}{2}\,.
\eea
Note that we use $\tilde{\omega}$ for the frequency, in order to distinguish it from the symplectic form.

\begin{figure}[ht!]
\centering
\includegraphics[width=12cm, angle = 0]{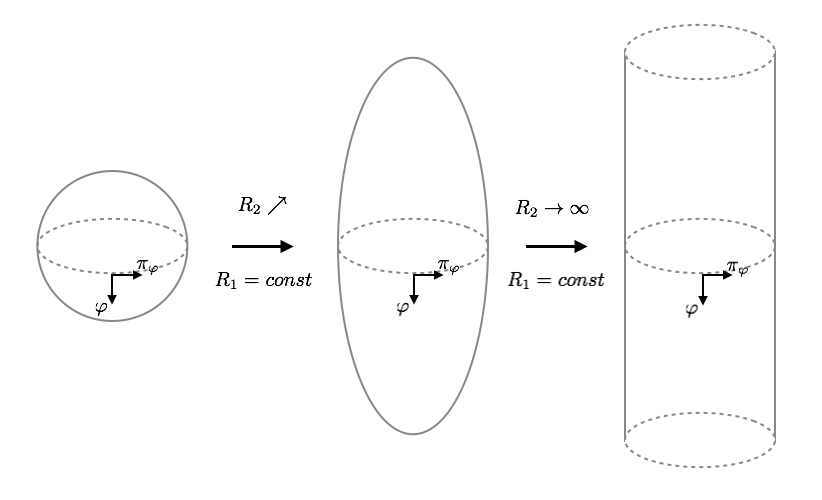}
\caption{Deformation of the original phase space $S^2$ to the cylindrical phase space 
of the polymerised dynamics.}
\label{Fig3}
\end{figure}

Let us now consider the case where we take the limit of $R_2 \rightarrow \infty$, whereby one recovers the standard Poisson brackets
\bea
\{\varphi, \pi_\varphi  \} = 1.
\eea
Taking this limit essentially means that the phase space is now not a spherical one anymore, but is rather bounded in one direction while being unbounded in another. Indeed, we go from $S^2 \rightarrow \mathbb{R} \times S^1$. With this limit, the spin components take the form
\bea
S_x &=& S\cos\(\frac{\pi_\varphi}{R_1}\)\,,\\
S_y &=& S\sin\(\frac{\pi_\varphi}{R_1}\)\,,\\
S_z &=& -S\(\frac{\varphi}{R_2}\)\,.
\eea
The first thing to ensure is that, in this limit, the Poisson bracket between, say, $S_y$ 
and $S_z$ is consistent with the symplectic form on the phase space. We get
\bea
\left\{\frac{\sin\(\frac{\pi_\varphi}{R_1}\)}{\frac{1}{R_1}}, -\varphi\right\} 
= \cos\(\frac{\pi_\varphi}{R_1}\)\,,
\eea
where we have again used $S=R_1R_2$. Let us now look at the form of the 
Hamiltonian (\ref{HamHO}) in this limit, which is 
\bea
H=\alpha\(1-\cos\(\frac{\pi_\varphi}{R_1}\)\) = 
\frac{1}{2m}    \frac{\sin^2 (\lambda  \pi_\varphi) }{\lambda^2}\,,
\eea
where $\lambda := \frac{1}{2R_1}$ denotes the scale of polymerisation. For $\lambda \rightarrow 0$, the kinetic term of the non-relativistic massive particle is recovered. The performed procedure of elongation of the initially spherical phase space into the final cylindrical phase space of the polymerised theory has beed depicted in Fig. \ref{Fig3}.

The above results should look very familiar to researchers working in loop quantum cosmology (LQC) and loop quantum gravity (LQG) since, what we obtained, is precisely the mentioned  ``polymerisation'' of the momentum. There also exists a sizable literature on the polymerised quantum mechanics of the harmonic oscillator. The polymerisation was also broadly studied in field theory \cite{Hossain:2009vd,Hossain:2010eb}. 

However, it is important to keep in mind that we did not introduce a modification of the form $\pi_\varphi \rightarrow \sin(\pi_\varphi)$. Rather, this latter comes out from keeping one of the phase space directions bounded while letting the other go to infinity. This is precisely what is done in the field space of LQC, where one regularises the curvature components in a way such that the connection components get replaced by bounded operators, coming from holonomies, whereas the triad components remain unbounded.

It is worth commenting on the limits we have hitherto taken in the above expressions. In order to derive the harmonic oscillator Lagrangian, in the leading order, we first took the limit of small angles for both $\varphi$ and $\pi_\varphi$. The next limit is essentially that we can still make the small angle approximation for the $\varphi$ variable but not for the conjugate momentum. Thus, we are not allowed to expand the trigonometric identities for $\pi_\varphi$.

It is also worth noticing that while Hamiltonian (\ref{HamHO}) reduces to the harmonic oscillator in the limit of low excitations, it does not lead to the polymerised harmonic oscillator in the $R_2 \rightarrow \infty$ limit. Recovering a Hamiltonian of the polymerised harmonic oscillator would require a modification of the original Hamiltonian (\ref{HamHO}). 

Another possibility one can derive from this observation is that necessarily the polymerisation of the configuration variables cannot be performed without modifying the conjugate variables as well, as considered e.g. in Refs. \cite{GPCamp,Hinter}. In these latter references, polymerisation schemes have been considered in which one modifies both the extrinsic curvature component and the conjugate flux variable, due to other considerations about solving the Hamiltonian constraint in LQG or avoiding quantum gravitational anomalies.

Actually, such possibility has already been hypothesised at the level of LQG in 2+1 dimensions \cite{Rovelli:2015fwa}. The standard LQG has partially compact phase space, which is  $SU(2) \times su(2) \ni (e^{\int A}, F[E])$ per link. A holonomy $e^{\int{A}}$ of the Ashtekar connection $A$ is the element of a compact group, while the fluxes of the densities triads $E$ belong to the the  $su(2)$ algebra. Recently, an alternative proposal in which the densities triad has been a subject of exponentiation, while the connection belongs to a flat part of the phase space, has been considered \cite{Bahr:2015bra}, namely $(F[A], e^{\int E}) \in \times su(2) \times  SU(2)$. A next step to perform would then amount to compactly account for both the phase space part of $A$ and $E$, such that $(e^{\int A}, e^{\int E}) \in SU(2)\times SU(2)$ would become a fully compact phase space. \\

In summary:

\begin{itemize}
\item {\bf Past}. LQG. $(e^{\int A}, E) \in  SU(2) \times su(2)$ $\rightarrow$ polymerization of $A$. 
\item {\bf Present}.  New LQG.  $(A, e^{\int E}) \in   \times su(2) \times  SU(2)$ $\rightarrow$ polymerization of $E$. 
\item {\bf Future}. Compact phase space LQG? $(e^{\int A}, e^{\int E}) \in SU(2)\times SU(2)$ $\rightarrow$ compact 
phase space quantum gravity based  fully on the spin phase space.
\end{itemize}

The idea of SFC we presented here focusing on the example of the scalar field seems to fit naturally into the picture of compact phase space version of LQG. The extension of the SFC to the realm of the gauge field theories is therefore desirable and will be discussed in our forthcoming articles. 

\section{Summary and outlook}\label{Sec_Summary}

We have shown that the relativistic Klein-Gordon field can be considered as an approximate description of the XXZ Heisenberg model for leading perturbations around the vacuum state and for the anisotropy parameter $\Delta \rightarrow 0$.  It is worth to notice that this limit is associated with the  XY model, describing a superconducting state of matter \cite{KT}. 

The deviation from the relativistic case, parametrised by the dimensionless parameter $\Delta$, affects the space-time symmetries, yielding features that might be confused with non-commutative Hopf algebras. Nonetheless, the deformed 
bi-algebroid we recovered has not a Hopf algebra structure. This shows that in the $\Delta\neq0$ case the Poincar\'e algebra is broken, a result that we was expected given the presence of a magnetic fields $\vec{B}$ that breaks isotropy. Indeed, the leading order interactions predicted by the theory break the Lorentz symmetries, as can be seen from the 
direct inspection of the co-algebraic structure. 

Furthermore, we have shown that the approach developed in this article,  which deals with an effective theory of real scalar field that is recovered in the continuum limit, can be reduced to the polymer quantisation in a limit when the spherical phase space is elongated in one direction. Such polymerisation of variables is considered as a simplified version of the quantisation procedure adopted in loop quantum gravity. The obtained results shed a new light on the possible relevance of the compact phase space field theories in the context quantum gravity (see also Ref. \cite{Trzesniewski:2017lpb}), while suggests that the fundamental theory to be quantized encodes irreducible representations of a compact 
Lie group $G$.  

There are various possible applications of the proposed framework. The most relevant one allows to interpret the fundamental scalar fields as a leading order excitation of some underlying spin systems. The nature of such hypothetical spin systems is, however, unknown at the moment. More direct applications include the construction of the condensed matter analog models of the field theoretical systems, providing the possibility of experimental tests of the quantization procedure adopted. Furthermore, since scalar fields are broadly considered in the description of the the evolution of the universe, application of the developed effective framework in cosmology seems to be natural \cite{Mielczarek:2017zoq} in this context.  Nonetheless, this extension also suggests that its regime of validity must be confined to effective theories, 
the fundamental one being associated to spin-chain systems, which can be recreated in analogues contexts in the laboratory.

The procedure we presented here still deserves generalisation to the other types of field, 
including spinor and gauge fields, and gravity. This issue will be addressed in our further studies. \\

\acknowledgments

A.M. wishes to acknowledge support by the Shanghai Municipality, through the grant No. KBH1512299, 
and by Fudan University, through the grant No. JJH1512105. J.M. has been supported by the Iuventus 
Plus grant 0302/IP3/2015/73 from the Polish Ministry of Science and Higher Education and National 
Science Centre (Poland) project DEC-2014/13/D/ST2/01895.

\end{document}